# A note on the ranking of earthquake forecasts

## G. Molchan

*Institute of Earthquake Prediction Theory and Mathematical Geophysics*

*84/32, Profsoyuznaya Str., Moscow,117997, Russian Federation.*

*E-mail: molchan@mitp.ru*

**Abstract.**

The ranking problem of earthquake forecasts is considered. We formulate simple statistical requirements to forecasting quality measure R and analyze some R-ranking methods on this basis, in particular, the pari-mutuel gambling method by Zechar&Zhuang (2014).

## 1. Introduction.

The task of earthquake forecasting is to predict a map $\{\lambda(\Delta_k)\}$ of target events rate in cells $\{\Delta_k\}$ of a region $G$ for a time period $\Delta T$. Reliable estimates of $\{\lambda(\Delta_k)\}$ are usually problematical. For this reason there are many approaches to the estimation of such maps that are not based on instrumental data alone. This circumstance is incorporated in the Regional Earthquake Likelihood Models (RELM) project (Schorlemmer et al., 2007), and this stimulated comparison and ranking of the normalized models

$$P^{(\alpha)} = \{\lambda^{(\alpha)}(\Delta_k)/\Lambda^{(\alpha)}\} = \{p^{(\alpha)}(\Delta_k)\}, \quad \Lambda^{(\alpha)} = \sum_k \lambda^{(\alpha)}(\Delta_k), \quad \alpha = 1 \div n; k = 1 \div K \quad (1)$$

depending on observations $\dot{N}_{\Delta T} = \{n(\Delta_k)/N\}$. Here, $n(\Delta)$ is the number of target events in a cell during a time $\Delta T$, and $N$ is the sum of these numbers. Independently, the comparison problem of the models (1) is highly important in seismic risk analysis.

The ranking problem is solved in the project by pairwise comparison of the competing models (Schorlemmer et al., 2007; Eberhard et al., 2012; Zechar et al., 2013). This approach is based on the classical theory of hypothesis testing, where it is always assumed that one of two competing hypotheses is true. Therefore the matrix of all pairwise comparisons among $\{P^{(\alpha)}, \alpha = 1,...,n > 2\}$ cannot provide a logically consistent ranking of the models (Rhoads et al., 2011; Molchan, 2011). New approaches arose as a result. Some of these are based on comparative and integrated analysis of maps (Clements et al., 2011; Taroni et al., 2014), while others try to rank models using a measure of forecasting quality, $R$ ( Zechar &Zhuang, 2014). The R-ranking method proposed in the last paper is purely intuitive. Our goal is to formulate desirable properties of $R$ at least on the stage of large $N$ and to analyze some R-ranking methods on this basis.

## 2. The basis for the R-ranking.

To be able to discuss forecasting quality measures, it is sufficient to consider the simplest seismicity model. We assume that $\{n(\Delta_k)\}$ come from a Poisson sequence of events with the rate $\lambda(\Delta_k)$ at $\Delta_k$. For fixed $N = \Sigma n(\Delta_k)$, the conditional distribution $\{n(\Delta_k)\}$ is a polynomial one, i.e.,

$$p(n(\Delta_1) = n_1,...,n(\Delta_K) = n_K) = N!/(n_1!..n_K!) p(\Delta_1)^{n_1} \cdot ... \cdot p(\Delta_K)^{n_K}. \qquad (2)$$

The problem of agreement of the model $P^{(\alpha)} = \{p^{(\alpha)}(\Delta_k)\}$ with the data $\dot{N}_{\Delta T} = \{n(\Delta_k)/N\}$ will be considered for large $N$. Under the conditions of stationarity, $\dot{N}_{\Delta T}$ converges to the exact distribution $P^{\bullet} = \{p^{\bullet}(\Delta_k)\}$ as $N \to \infty$.

## 3. The quality measure of the model, $R$.

The quality measure of the model, $R$, can be treated as a continuous functional of three components: the empirical distribution $\dot{N}_{\Delta T} = \{n(\Delta_k)/N\}$, the model $P = \{p(\Delta_k)\}$, and a reference model $P^0 = \{p^0(\Delta_k)\}$, i.e., $R = R\{\dot{N}_{\Delta T}, P, P^0\}$. The model $P^0$ is a point of reference; it is therefore desirable that its choice should not substantially affect the ranking of the competing models (the **0-requirement** on $R$).

Suppose that one of the competing models is explicit, $P^{\bullet} = \{p^{\bullet}(\Delta_k)\}$; then for large $N$ this model must almost surely get the highest rating. In the limit, $N \to \infty$, this entails the **1-requirement** on $R$:

$$R\{P^{\bullet}, P^{\bullet}, P^0\} > R\{P^{\bullet}, P, P^0\}, \quad P^{\bullet} \neq P. \tag{3}$$

In the above, we have replaced the first argument $\dot{N}_{\Delta T}$ with its limiting value $P^{\bullet}$. Requirement (3) also presupposes that, if (3) turns into equality then $P$ is identical with the exact distribution. This important requirement can be extended.

Suppose $R\{\dot{N}_{\Delta T}, P, P^0\}$ is used as a statistic for testing the model $P$. Assuming $P$ is true, $R$ must have the limit $R\{P, P, P^0\}$ as $N \to \infty$; actually the limit is $R\{P^{\bullet}, P, P^0\}$. The $R$ test cannot reject the wrong $P$ model, if these limits are identical (Molchan, 2011). For this reason it would be *desirable* to have the following

**requirement 2:**

$$\text{if} \quad R\{P, P, P^0\} = R\{P^{\bullet}, P, P^0\} \quad \text{then} \quad P^{\bullet} = P. \tag{4}$$

In mathematical statistics this requirement is considered as the *consistency* property of the statistical test $R$ (Borovkov, 1984).

Obviously, any metric $\rho\{P_1, P_2\}$ on the space of distributions satisfies requirements **1** and **2** and can be used to rank models by their distances from the empirical distribution $\dot{N}_{\Delta T} = \{n(\Delta_k)/N\}$. But any choice of the metric is dictated by additional requirements. To give an example, divide the seismic region $G$ into zones $J^{(r)}$ of different degrees of importance for some user, choose a suitable metric $\rho(P_1.P_2)$ and construct a new metric:

$\rho_w(P_1.P_2) = \sum_r \rho_r(P_1, P_2) w_r$, where $\rho_r$ is a restriction of $\rho(P_1.P_2)$ to $J^{(r)}$, while the weight $w_r$ increases with the importance of $J^{(r)}$. In that case the principle of ranking the distributions $P^{(\alpha)}$ according to how near they are to $\dot{N}_{\Delta T}$ will be based on the principle of usefulness, which remains a subjective one.

Consider some known examples of $R$.

## 4. Information score $R$.

Let $L(N_{\Delta T}|P)$ be the log-likelihood of $N_{\Delta T} = \{n(\Delta_k)\}$ relative to the model $P$. We define $R\{\dot{N}_{\Delta T}, P, P^0\} = [L(N_{\Delta T}|P) - L(N_{\Delta T}|P^0)]/N$. Due to (2), one has

$$R\{\dot{N}_{\Delta T}, P, P^0\} = \sum_k \dot{n}_k \log(p_k / p_k^0), \tag{5}$$

where $\dot{n}_k = n(\Delta_k)/N$. The RELM-project uses $R\{\dot{N}_{\Delta T}, P^{(\alpha)}, P^{(\beta)}\}$ for pairwise comparison among the models.

It is easy to see that the information score $R$ satisfies requirement **1**. Indeed, by (5), one has

$$R\{P^\bullet, P^\bullet, P^0\} - R\{P^\bullet, P, P^0\} = \sum_k p_k^\bullet \log(p_k^\bullet / p_k) = I(P^\bullet, P) \tag{6}$$

where $I(P_1, P_2)$ is the information "distance" between $P_1$ and $P_2$. It is well known that $I(P_1, P_2)$ is nonnegative, and is zero only when $P_1 = P_2$. For this reason (6) yields (3).

To verify requirement **2**, let us consider equation (4). Due to (5), this equation looks as follows

$$\sum_k (p_k^\bullet - p_k)\log(p_k / p_k^0) = 0. \tag{7}$$

Considering $\{p_k^\bullet\}$ as variables, (7) is a hyperplane that passes through a point $P$ of the simplex $S: \sum_k p_k^\bullet = 1$, $0 \leq p_k^\bullet \leq 1$. The intersection of these geometrical objects defines the set of solutions $\{P^\bullet\}$ of equation (7). The intersection is obviously a simplex $S' \subset S$ of dimension $K-2$, provided $P$ is an inner point on S. For this reason requirement **2** is violated.

## 5. The pari-mutuel gambling score $R$.

Let us approximate the information score (5) using the relation $\log(x) \approx x - 1$ for $x \sim 1$. We get a new version of the $R$ score, viz.,

$$R_1\{\dot{N}_{\Delta T}, P, P^0\} = \sum_k \dot{n}_k (p_k / p_k^0 - 1) = \sum_k (\dot{n}_k - p_k^0)(p_k - p_k^0)/p_k^0 \tag{8}$$

Using formal analogies with gaming, Zechar&Zhuang (2014) came to a symmetric version of (8), namely

$$R\{\dot{N}_{\Delta T}, P, P^0\} = \sum_k (\dot{n}_k - p_k^0)(p_k - p_k^0)/[p_k^0(1-p_k^0)]$$

$$= R_1\{\dot{N}_{\Delta T}, P, P^0\} + R_1\{\overline{\dot{N}}_{\Delta T}, \overline{P}, \overline{P}^0\}(K\text{-}1) \quad, \tag{9}$$

where $\overline{P} = \{(1-p_k)/(K-1)\}$, and $K$ is the number of cells. In addition, the authors choose $P_0$ as follows:

$$P_0 = \sum_\alpha w_\alpha P^{(\alpha)}, \qquad \sum_\alpha w_\alpha = 1 \quad w_\alpha \geq 0 \tag{10}$$

where $w_\alpha = 1/n$ and $n$ is the number of competing models. The $R$ score (9,10) is the basis of the pari-mutuel gambling method in the merit ranking of forecasts.

Practically in the same form, the *R* score was used by Zechar&Zhuang (2010) for testing of binary forecasts. However, for this purpose the *R* score (9) was potentially unstable ( Molchan & Romashkova, 2011) . Now we are going to check whether the *R* function (9,10) can be applied to probabilistic forecasts.

Inequality (3) combined with (9) looks as follows:

$$R\{P^\bullet, P^\bullet, P^0\} - R\{P^\bullet, P, P^0\} = \sum_k (p_k^\bullet - p_k^0)(p_k^\bullet - p_k)/[p_k^0(1-p_k^0)] > 0, \quad P^\bullet \neq P. \quad (11)$$

At first we consider the case when $P^0$ is independent of $P$. Then the right-hand side of (11), considered as function $\Delta R$ of variables $\{p_k\}$, is linear. All distributions $P$ form the simplex $S$: $\sum_k p_k = 1$, $0 \leq p_k \leq 1$. Obviously, the hyperplane $\Delta R = 0$ has a common point $P^\bullet$ with $S$. For this reason the intersection $\{\Delta R = 0\} \cap S$ is a simplex again. In the generic situation, the dimension of the intersection is K-2. That means that the right-hand side of (11) is an alternating function on $S$. Therefore, in general, the requirement **1** for $R$ is not satisfied.

Let us consider now the original case in which $P^0$ is given by (10). We fix the distribution $Q = \sum_{\alpha=1}^{n-1} w_\alpha P^{(\alpha)}/(1-w_n) := \{q_k\}$ and consider $\Delta R$ as a function of $P = \{p_k\} = P^{(n)}$ on the simplex $S$: $\sum_k p_k = 1$, $0 \leq p_k \leq 1$. One has $P^0 = (1-w_n)Q + w_n P$ and

$$\Delta R(P) = \sum_k (p_k^\bullet - (1-w_n)q_k - w_n p_k)(p_k^\bullet - p_k)/[p_k^0(1-p_k^0)] \quad (12)$$

Obviously, $\Delta R = 0$ at point $P^\bullet$ on $S$. If $\Delta R$ has fix sign in a vicinity of $P^\bullet$ on $S$, then $P^\bullet$ have to be the stationary point of $\Delta R(P)$ on $S$, i.e.,

$$(\partial/\partial p_k)\Delta R(P)\big|_{P=P^\bullet} = \lambda, \quad k = 1,...,K, \quad (13)$$

Here the unknown constant $\lambda$ is a result of the condition: $\sum_k p_k = 1$. Using (13) , one has

$$-(1-w_n)(p_k^\bullet - q_k) = \lambda \tilde{p}_k^0(1-\tilde{p}_k^0), \qquad k=1,\ldots,K, \tag{14}$$

where $\tilde{p}_k^0 = (1-w_n)q_k + w_n p_k^\bullet$. By summing (14) over $k$, we conclude that

$$0 = \lambda \sum_k q_k(1-q_k) \quad \text{and } \lambda = 0. \tag{15}$$

Using (14), we get $P^\bullet = Q$. In the generic situation $P^\bullet \neq Q$. Hence, $\Delta R(P)$ is an alternating function on $S$. In other words, there are a lot of models $P$ such that $P^\bullet \prec P$ relative to the pari-mutuel gambling score under the condition $N \gg 1$.

Let as give an explicit example. For simplicity we will consider infinitely small cells $\Delta_k$. In this case

$$\Delta R(P) = \int_G (p^\bullet(g) - p(g))p^\bullet(g)/p^0(g)dg, \tag{16}$$

where $p^\bullet(g), p(g), p^0(g)$ are densities of the distributions $P^\bullet, P, P^0$ respectively.

Let $G = (0,1)$, $p^\bullet(g) = 1$; the number of competing hypotheses is three, n=3; $p^0(g) = (p^{(1)} + p^{(2)} + p^{(3)})/3$, where $p^{(1)}(g) = p^\bullet(g) = 1$; $p^{(2)}(g) = \varepsilon^{-1}[0 < g < \varepsilon]$ and $p^{(3)}(g) = \delta^{-1}[1-\delta < g < 1]$, $\varepsilon + \delta < 1$. Here $[A]$ is the 0-1 logical function.

Setting $P = P^{(2)}$, we can continue (16):

$$(16) = 3\int_0^1 (1-p^{(2)}(g))/(1+p^{(2)}(g)+p^{(3)}(g))dg = 3(1-\varepsilon(1+2\delta))(1+\varepsilon)^{-1}(1+\delta)^{-1}. \tag{17}$$

The right-hand side of (17) is obviously negative as soon as $\varepsilon(1+2\delta) > 1$; in particular, it is true when $\varepsilon = 0.7$ and $\delta = 0.25$. We get the undesirable relation $P^{(1)} = P^\bullet \prec P^{(2)}$.

Finally note, that the information analogue of the pari-mutuel gambling score, namely,

$$R\{\dot{N}_{\Delta T}, P, P_0\} = \sum_k \dot{n}_k \log(p_k / p_{k0}) + \sum_k \bar{\dot{n}}_k \log(\bar{p}_k / \bar{p}_{k0}) \quad (14)$$

with the reference distribution (10) is in agreement with the 1-requirement and therefore looks preferable.

## 6. Conclusion

Maps of the seismicity rate for the current time can be treated as earthquake forecasting. There are many objective tests to select maps so they are in agreement with the data. However, attempts at ranking the selected models using a quality measure simplify the problem unjustifiably. We have formulated simple requirements to forecasting quality measure and shown that the parimutuel gambling method by Zechar &Zhuang, (2014) does not necessarily rank the exact seismicity model correctly, even with extensive data. In general, the integrated approaches look preferable for finding the "best" forecast among the competing models.